\documentclass[aps,twocolumn]{revtex4}
\usepackage[english]{babel}
\usepackage{amsfonts}
\usepackage{amsmath}
\usepackage{amssymb}
\usepackage{latexsym}
\usepackage{hyperref}
\usepackage{grffile}
\usepackage{graphicx}
\usepackage{epstopdf}

\def\cacu{CaCu$_3$Fe$_2$Re$_2$O$_{12}$}
\def\cucu{CuCu$_3$Fe$_2$Re$_2$O$_{12}$}

\begin{document}
\title{Highly correlated electronic state in a ferrimagnetic quadruple perovskite CuCu$_3$Fe$_2$Re$_2$O$_{12}$}

\author{
A.\,I.\, Poteryaev$^{a,b}$, 
 Z.\,V.\, Pchelkina$^{a,b}$,
S.\,V.\, Streltsov$^{a,b}$,
Y.\, Long$^c$, 
V.\,Yu.\, Irkhin$^{a}$}
  


\affiliation{$^a$Institute of Metal Physics, S. Kovalevskaya Street 18, 620108 Ekaterinburg, Russia\\~\\
$^b$Department of Theoretical Physics and Applied Mathematics, Ural Federal University,
Mira St. 19, 620002 Ekaterinburg, Russia\\~\\
$^c$Beijing National Laboratory for Condensed Matter Physics, Institute of Physics, Chinese Academy of Sciences, Beijing 100190, China\\~\\
}

\begin{abstract}
Recently synthesized  quadruple perovskite CuCu$_3$Fe$_2$Re$_2$O$_{12}$ possesses strong ferromagnetism and unusual electron properties, including enhanced electronic specific heat. Application of the first principles electronic structure approaches unambiguously shows importance of the many-body effects in this compound. While CuCu$_3$Fe$_2$Re$_2$O$_{12}$ is half-metallic ferrimagnet in the DFT+U method, in the density functional theory (DFT) combined with the dynamical mean-field theory (DMFT) it appears to be a metal. Strong  correlations lead to a renormalization of electronic spectrum and formation of incoherent states close to the Fermi level. Electronic specific heat and magnetic properties obtained in the DFT+DMFT approach are in better agreement with available experimental data than derived by other band structure techniques.
\end{abstract}


\maketitle

\section{Introduction} 
A number of quadruple perovskites $A$Cu$_3$B$_2 B'_2$O$_{12}$  possess very interesting electron and magnetic properties. 
Some of them, e.g., CaCu$_3$Fe$_2$Re$_2$O$_{12}$ \cite{Chen}, NaCu$_3$Fe$_2$Os$_2$O$_{12}$ \cite{Wang1}, NaCu$_3$Fe$_2$Re$_2$O$_{12}$ \cite{Wang2} and LaCu$_3$Fe$_2$Re$_2$O$_{12}$  \cite{Zhehong} possess ferromagnetism with large magnetic moment and high Curie temperatures, and are supposed to be half-metallic ferromagnets (HMF's) or ferrimagnets.

On the other hand, there exist a number of paramagnetic compounds 
with  high values of $\gamma T$-linear electronic specific heat. For CaCu$_3$Cu$_2$Ir$_2$O$_{12}$,  $\gamma$ = 211 mJ/mol~K$^2$ \cite{Zhao}; for CaCu$_3$Ir$_4$O$_{12}$  experimental $\gamma$ = 173 mJ/mol~K$^2$, but LDA calculations give a much smaller value 17 mJ/mol~K$^2$, demonstrating that the electronic specific heat is strongly enhanced by electronic correlations, possibly of Kondo origin \cite{Cheng}. Interestingly, $\gamma$ in Ca$_{1-x}$Y$_x$Cu$_3$Co$_4$O$_{12}$ decreases from 157 mJ/mol K$^2$ for $x=0$ to 47 mJ/mol K$^2$ for $x=1$ \cite{Yamada}. For the system CaCu$_3$Ru$_{4-x}$Fe$_x$O$_{12}$,  the linear electronic specific heat coefficient increases with  Fe doping from 90 mJ/mol~K$^2$ for $x=0$ to 271 mJ/mol~K$^2$ for $x=0.2$, indications of magnetic order being observed for  all the Fe doped samples \cite{Wang}.

Strongly ferromagnetic quadruple perovskites can  have enhanced $\gamma$ values (e.g., 95~mJ/mol K$^2$ for LaCu$_3$Co$_2$Re$_2$O$_{12}$ \cite{Liu} and about  35~mJ/mol K$^2$ for CaCu$_3$Fe$_2$Re$_2$O$_{12}$ \cite{Youwen}). In CuCu$_3$Fe$_2$Re$_2$O$_{12}$ $\gamma = $ 62 mJ/mol~K$^2$ \cite{Youwen}.
Besides that, CuCu$_3$Fe$_2$Re$_2$O$_{12}$ demonstrates a strongly anharmonic lattice specific heat, which is described in terms of rattling motion of Cu ion recently found in Ref.~\cite{Pchelkina} (a feature which is absent for CaCu$_3$Fe$_2$Re$_2$O$_{12}$).  At the same time, the saturation magnetic moment ($\sim$5~$\mu_B$) and Curie temperature $T_C =190$ K are considerably lower than those for CaCu$_3$Fe$_2$Re$_2$O$_{12}$ with large saturated magnetization of 8.7~$\mu_B$ and   $T_C =560$~K \cite{Chen}. 

These properties of CuCu$_3$Fe$_2$Re$_2$O$_{12}$ are hardly compatible with half-metallic ferromagnetism.
Indeed, HMF's, being typically strong itinerant ferromagnets, possess in one of spin channels an energy gap  which  can have different nature, so that their magnetic moment tends to a maximum possible value. The gap can originate either from hybridization or from the Hubbard splitting, as in strongly correlated systems \cite{Katsnelson}. 
In the HMF state, usual spin-fluctuation mechanisms of effective mass enhancement do not work since spin-flip processes are forbidden (although some mechanisms owing to incoherent states can give a contribution to specific heat \cite{Katsnelson1}). Thus enhanced $\gamma$ values in ferromagnetic quadruple perovskites provide a challenge, and the problem of describing their ground state occurs.

In the present paper we perform first-principle investigations of electronic structure and magnetic state in the quadruple perovskite CuCu$_3$Fe$_2$Re$_2$O$_{12}$. To this aim the conventional density functional theory (DFT) is used for understanding basic properties; the DFT+U method~\cite{LDAU} allows one to take into account the static electron-electron correlations; the state of the art DFT+DMFT approach~\cite{DFTDMFT,DFTDMFT_Licht}, that is a combination of the density functional theory and dynamical mean-field theory~\cite{DMFT}, treats correlation effects in dynamical way and is employed to study correlation effects in more details.  Based on the results obtained, we report evidences of considerable electron correlations in \cucu.

\section{ Results}  
\begin{figure}[t!]
\centering
\includegraphics[width=0.49\columnwidth, trim={17cm 15cm 105cm 3cm}, clip]{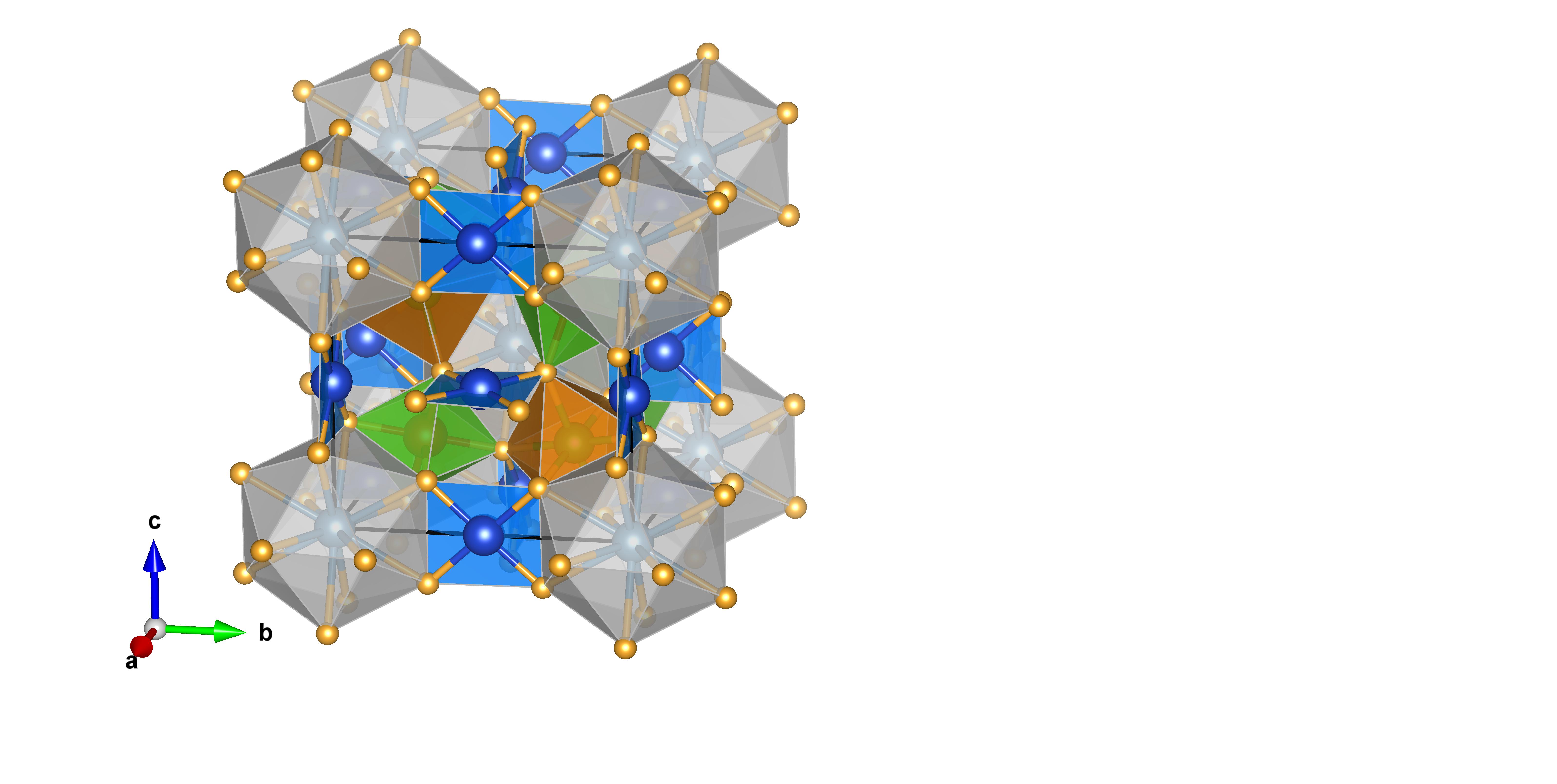}
\includegraphics[width=0.49\columnwidth, trim={17cm 15cm 105cm 3cm}, clip]{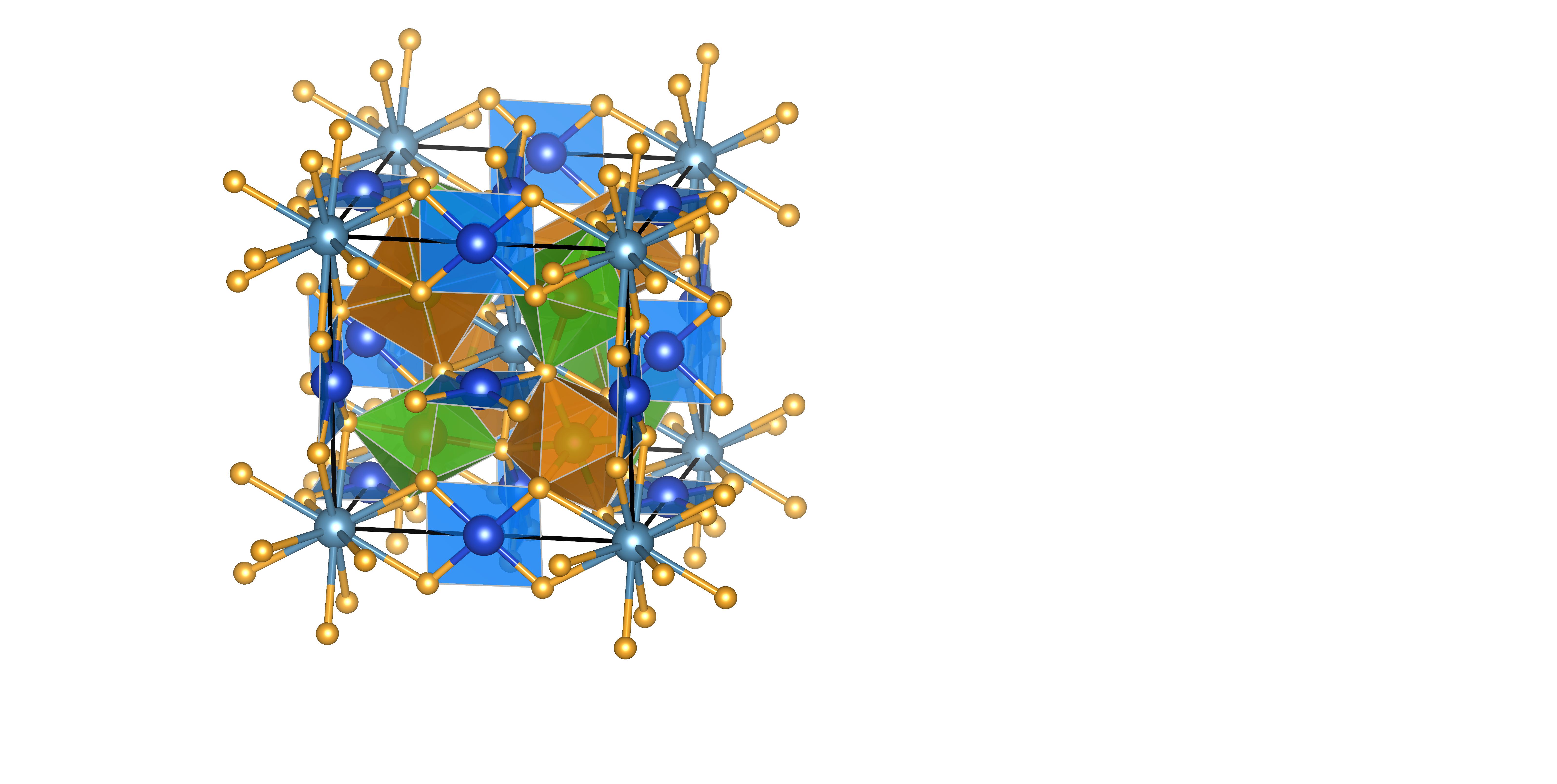}
\caption{Fig. 1. Left panel illustrates crystal structure of \cucu~ with different type of oxygen (gold color) polyhedra surrounding transition metal sites. Right panel shows the same structure with an eliminated  icosahedron around Cu$_1$ site (light blue color) and  square plaquette around Cu$_2$ (blue). The network of tilted green and orange octahedra becomes visible around Fe and Re, correspondingly.}
\label{fig:CaCuFeRe_structure}
\end{figure}

The crystal structure for \cucu~ with the $Pn$-3 space group (number 201) is shown in Fig. \ref{fig:CaCuFeRe_structure}. The Cu atoms occupy two crystallographically nonequivalent positions. The first type of Cu (Cu$_1$=$A$ specie) are located at the corners of the cubic primitive cell and they are surrounded by the oxygen icosahedra shown in grey color (left panel of Fig.~\ref{fig:CaCuFeRe_structure}). 
Second type of Cu ions (Cu$_2$=$A'$ specie) have square plaquette oxygen environment in which degeneracy of Cu 3$d$ shell is completely removed. The transition metal ions Fe and Re occupy $B$ and $B'$ positions, correspondingly, and form the network of corner-shared octahedra colored in green and orange in the right panel of Fig.~\ref{fig:CaCuFeRe_structure}. The $d$ bands of these transition metal ions are first split by the ligand field into $t_{2g}$ and $e_g^{\sigma}$ subbands. Both FeO$_6$ and ReO$_3$ are described by the $C_{3i}$ point group, which is lower than cubic $O_h$ symmetry. This results in additional splitting of $t_{2g}$ orbitals onto $a_{1g}$ and $e_g^{\pi}$ states.

\begin{table}
  \centering
  \caption{Table 1. {Partial $d$ orbital state} occupations, magnetic moments, and instant squared magnetic moments for transition metal ions obtained using different approaches. The superscripts encode various electronic structure approximations: DFT is the non-magnetic calculations, sp-DFT is spin-polarized version of above, DFT+U stands for the combination of the DFT and simplified static Hubbard correction to the band structure, and DMFT is a combination of the density functional theory and dynamical mean-field theory. {Since the Cu$_1$ 3$d$ states are completely occupied and regarded as uncorrelated in the DFT+DMFT method, the corresponding values are absent.}
}
  \label{tab:data}
  \begin{tabular}{l|cccc}
    \hline\hline
                     & Cu$_1$ & Cu$_2$  &  Fe    &    Re     \\  \hline
    $n^{DFT}$        & 9.44   &  9.179  & 6.183  &  4.582    \\  \hline
$n^{sp-DFT}$     & 9.433  &  9.187  & 5.962  &  4.558    \\ 
$m^{sp-DFT}$     & 0.024  &  0.068  & 3.659  & -0.111    \\  \hline
$n^{DFT+U}$       & 9.616  &  9.275  & 5.875  &  4.516    \\ 
$m^{DFT+U}$       & 0.004  &  0.524  & 4.062  & -0.859    \\  \hline
$n^{DMFT}$           &  -     &  8.893  & 5.381  &  3.907    \\ 
$m^{DMFT}$           &  -     & -0.011  & 4.276  & -0.527    \\
$\sqrt{\langle m_z^2 \rangle}$ & - & 0.982 & 4.434 & 1.343   \\  \hline\hline
  \end{tabular}
\end{table}

\begin{figure}
  \centering
  \includegraphics[width=0.95\columnwidth]{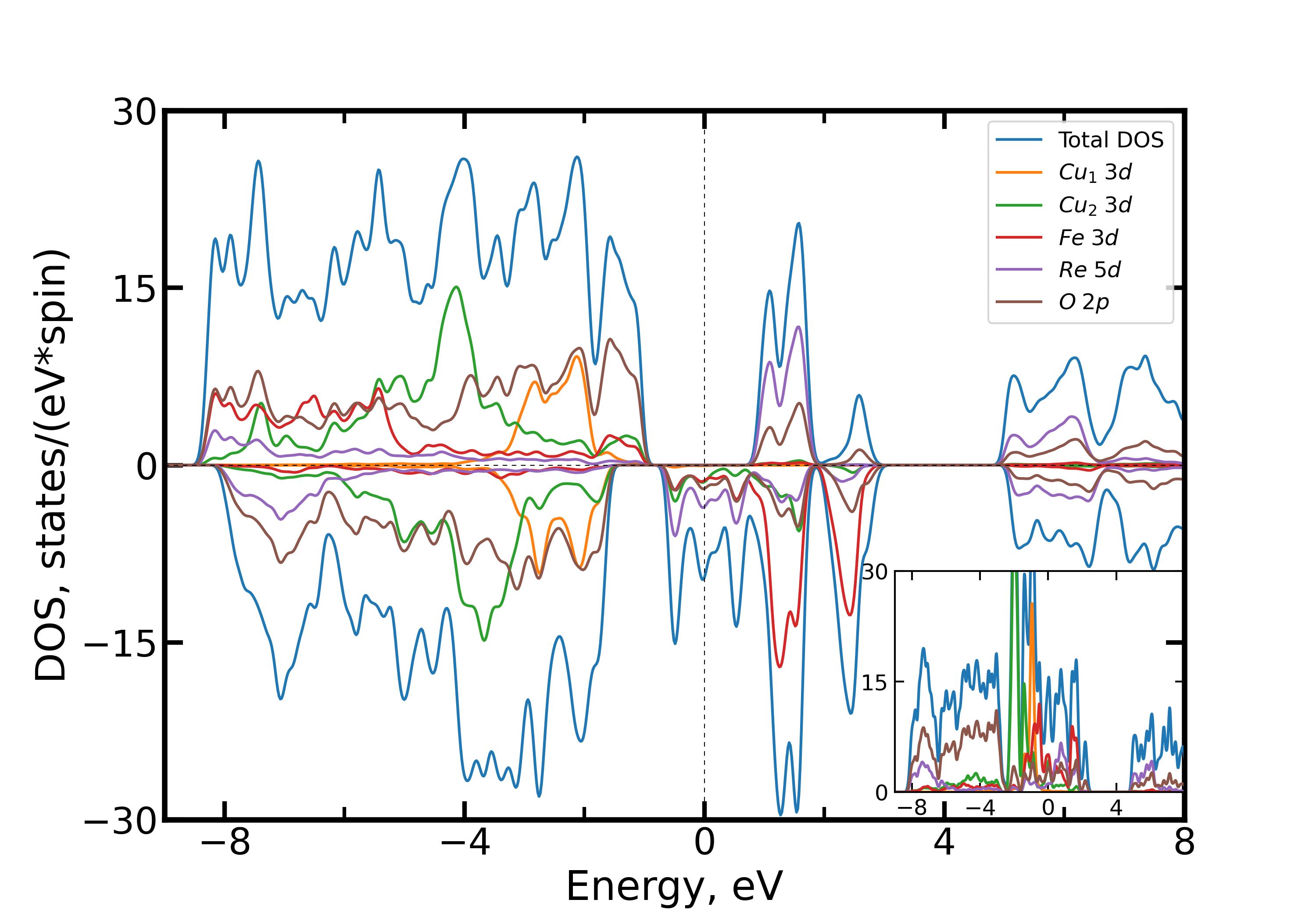}
  \caption{Fig. 2. DFT+U total and partial density of states (DOS) for \cucu. Cu$_1$ is in ligand icosahedron, while Cu$_2$  has square plaquette surrounding. Inset shows non-magnetic DFT DOS using the same color coding. { Since spin up DOS at the Fermi level is absent, but Fe and Re magnetic moments have opposite direction, the system is a half-metallic ferrimagnet. }
}
  \label{fig:CuCuFeRe_LDA_DOS}
\end{figure}

Fig.~\ref{fig:CuCuFeRe_LDA_DOS} shows the total and partial density of states (DOS) obtained in DFT+U and DFT approaches (inset of figure). The VASP package was used for this band structure calculations~\cite{VASP} in conjunction with the PBE exchange-correlation functional~\cite{PBE}. In all calculations we used the energy cutoff of 500~eV and the reciprocal space division of 8$\times$8$\times$8 {\bf k}-points. For transition metal ions, the following parameters of screened Coulomb interaction and Hund's exchange are used: $U_{Fe} =$ 4~eV, $J_{Fe} =$ 0.9~eV, $U_{Re} =$ 2~eV, $J_{Re} =$ 0.5~eV, $U_{Cu}=$ 7~eV, $J_{Cu}=$ 0.9~eV. 
{ These values of interactions are employed for conventional parameterization of the orbital-dependent Coulomb interaction matrix~\cite{LDAU},  being in a good agreement with reported data for transition metal compounds~\cite{BiFe,LaOFeAs,Zakharov}. 
}
Both types of the DFT+U scheme~\cite{Licht,Dudarev} available in VASP were used, and they provided very similar results with data shown for Dudarev's type of DFT+U~\cite{Dudarev}.

One can clearly see that according to DFT+U calculations \cucu~ is a half-metallic ferrimagnet with the gap in majority spin channel $\Delta_g^{\uparrow}$=1.88~eV, and density of states at the Fermi level for minority spin, $N_{\downarrow}(E_F)$ = 6.21~states/(eV f.u.). These are primarily $d$ states of Re, Fe, and Cu$_2$ ions with admixture of O $2p$ due to a strong hybridization. The 3$d$ states of Cu$_1$ ions, that sit inside icosahedra (orange color in Fig.~\ref{fig:CuCuFeRe_LDA_DOS}) are completely occupied, placed at about -2.5~eV and do not contribute to  the density at the Fermi level. The plaquette Cu$_2$ 3$d$ bands (green) lie below its Cu$_1$ counterpart. The majority spin states are completely occupied but minority spin states cross the Fermi level, which results in a small magnetic moment, $m_{Cu_2}$=0.52~$\mu_B$. The Fe 3$d^{\uparrow}$ band (red) is strongly hybridized with O 2$p$ states (brown) and  located  from -8 to -1~eV. The Fe 3$d^{\downarrow}$ band is almost empty that gives the magnetic moment $m_{Fe}$~= 4.06~$\mu_B$. The band structure of Re 3$d$ states (purple color in Fig.~\ref{fig:CuCuFeRe_LDA_DOS}) is more interesting: The majority spin  is almost empty with $t_{2g}$ and $e_g^{\sigma}$ states located from 1 to 2~eV and from 5 to 6.5~eV, respectively. The minority spin  is partially occupied and crosses the Fermi level. This leads to magnetic moment $m_{Re}$ = -0.86~$\mu_B$, which is opposite  to the magnetic moments directions of Fe and Cu (most probably due to antiferromagnetic superexchange). Hence, the overall ferrimagnetic configuration is Cu$^{\uparrow}$Fe$^{\uparrow}$Re$^{\downarrow}$, which is compatible with the result for \cacu~\cite{Chen}.  The total magnetization including not only contributions from atomic spheres is close to 10~$\mu_B$/f.u. { The  nearly integer value of magnetic moment, being determined by the total valence electron number according to the Slater-Pauling rule, is a typical situation for half-metallic magnetism.}

The low-energy DFT Hamiltonian suitable for the DMFT calculations can be built in conventional for transition metal compounds way. Namely, all bands that cross the Fermi level plus oxygen bands strongly hybridized with them  are included into the projected Hamiltonian. In case of \cucu, these are Cu 3$d$, Fe 3$d$, Re 5$t_{2g}$, and O 2$p$ states. Since empty Re $e_g^{\sigma}$ bands are located at 5 eV above the Fermi level, and they are well separated by the gap from the low-lying states, we excluded them from consideration to reduce computational costs. The non-magnetic band structure, which is a starting point for our DFT+DMFT calculations, is presented on the inset of Fig.~\ref{fig:CuCuFeRe_LDA_DOS}.  We used Wannier projection to obtain small Hamiltonian for DMFT calculations \cite{proj}. 
The completely occupied states of O 2$p$ and Cu$_1$  3$d$ are regarded as uncorrelated in our DFT+DMFT treatment. 
We used the same set of the interaction parameters as earlier in our DFT+U calculations. 
{To account for strong electron correlations in the DFT+DMFT framework, the AMULET package was employed~\cite{amulet}. The effective impurity problem arising in DMFT was solved with the continuous time quantum Monte Carlo method~\cite{Rubtsov2004,Gull2011} at the temperature $T=$ 200~K.}

\begin{figure}
  \centering
  \includegraphics[width=0.95\columnwidth]{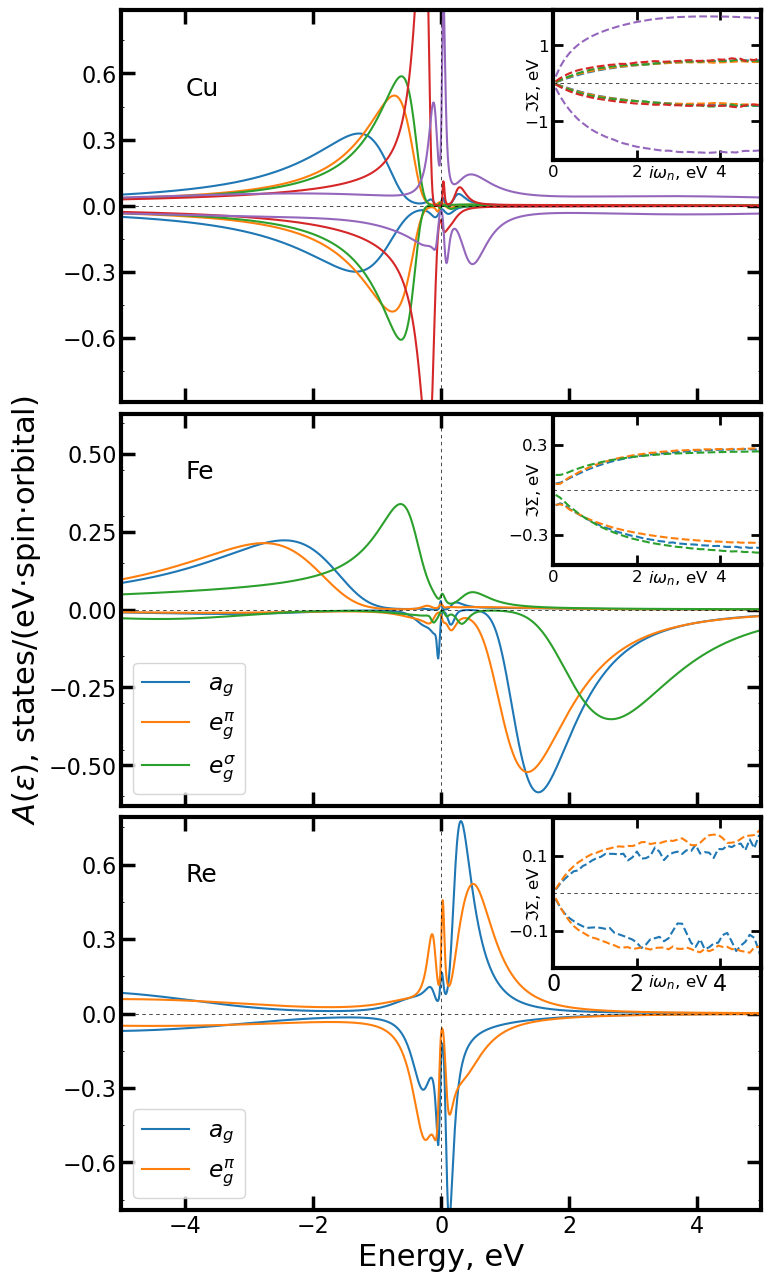}
  \caption{Fig. 3. DFT+DMFT spectral functions of \cucu. Various panels show the spin and orbital resolved spectral functions for Cu$_2$ (top), Fe (middle) and Re (bottom).  Cu$_2$  has square plaquette surrounding. Insets of corresponding panel show imaginary part of self-energies using the same color coding.}
  \label{fig:CuCuFeRe_DMFT_DOS_Sigma}
\end{figure}

Spin and orbital resolved spectral functions for correlated Cu, Fe, and Re states are presented in Fig. \ref{fig:CuCuFeRe_DMFT_DOS_Sigma}  (from top to bottom panel). The insets of this figure show the corresponding imaginary parts of self-energy $\Sigma(i\omega_n)$. Negative (positive) imaginary part of  $\Sigma(i\omega_n)$ corresponds to spin majority (minority). One can clearly see that all states are metallic for both spin projections. Therefore, \cucu~ is not half-metallic ferrimagnet in the DFT+DMFT approach. Cu$_2$ 3$d$ states (in plaquettes)  all except one are occupied. The partially occupied orbital has a pseudogap-like behavior near the Fermi level with the Fermi-liquid-like type of the imaginary part of the self-energy (purple color). The quasiparticle residue, $Z^{-1}=1-\partial\Im\Sigma/\partial\omega_n$ ($\omega_n$ is the Matsubara frequency), for this orbital is about 0.3, while for the rest of orbitals this value is close 0.7. 
This indicates a correlation enhancement of the electronic specific heat.

Ordered magnetic moment for Cu$_2$ is strongly reduced in the DFT+DMFT approach down to  -0.01~$\mu_B$ at $T=$ 200~K with the direction of the moment opposite to DFT+U one. At the same time, the instant squared magnetic moment, $\langle m_z^2 \rangle_{Cu_2}$~= 0.97~$\mu_B^2$, indicates that the Cu$_2$ 3$d$ states are in $d^9$ configuration with localized magnetic moment and ordered magnetic moment vanishing due to thermal fluctuations, which are absent in the DFT+U method. The majority spin Fe 3$d$ states are almost completely occupied as in DFT+U approach, but the gap in this spin channel is closed, and even more, there occurs a narrow quasiparticle-like peak at the Fermi level. The minority spin states are almost empty and have a pseudogap. The Fe 3$d$ ordered magnetic moment is 4.28~$\mu_B$, which is a bit larger than in DFT+U method. The instant squared magnetic moment of iron, $\langle m_z^2 \rangle_{Fe}~=$ 19.66~$\mu_B^2$, coincides nearly perfectly with the square of the ordered magnetic moment suggesting the absence of the longitudinal fluctuations.

The imaginary parts of the self-energies for both spin directions are incoherent with the divergent behavior at small $\omega_n$ (see inset of Fig.~\ref{fig:CuCuFeRe_DMFT_DOS_Sigma}). At the same time, this divergence is not strong enough to produce an insulating state. The Re $t_{2g}$ states have a narrow peak just above the Fermi level for the majority spin direction and a pseudogap in minority spin channel. The imaginary parts of self-energies go to zero at small $\omega_n$, and mean value for $Z_{Re}$ makes up about 0.83, which suggests moderate electronic correlations for this ion. Ordered magnetic moment on Re is reduced with respect to DFT+U counterpart and equals to $-0.53 \mu_B$. 
The total magnetic moment for \cucu~ obtained in the DFT+DMFT approach is 7.63~$\mu_B$ per f.u., which is strongly reduced with respect to the DFT+U value.

The value of the coefficient in the $T$-linear electronic specific heat, $\gamma$, calculated in the non-magnetic DFT, is determined by DOS at the Fermi level. From the inset of the Fig.~\ref{fig:CuCuFeRe_LDA_DOS} on see that there is a substantial peak at the Fermi level that results in $\gamma_{DFT}$~=~56 mJ/(mol K$^2$), which is $\sim$10\% 
smaller than the experimental value of $\gamma_{exp}$~=~62 mJ/(mol K$^2$). Nevertheless, this small discrepancy cannot be viewed satisfactory as \cucu~ is a magnetic compound. The large peak at the Fermi level in the non-magnetic solution leads to a magnetic instability.
This spin-polarized (sp) solution is metallic with a much smaller DOS at the Fermi level, that results in $\gamma_{sp-DFT}$~=~37 mJ/(mol K$^2$), which is smaller than its non-magnetic DFT counterpart. These inconsistencies suggest importance of  strong electron correlations in our system. 

Unfortunately, an attempt to account for strong electron-electron interaction in the DFT+U framework worsens the situation. In the above approach (see Fig.~\ref{fig:CuCuFeRe_LDA_DOS}) the obtained solution is a half-metallic ferrimagnet with a gap in one spin channel. This reduces drastically the linear coefficient to the value of $\gamma_{DFT+U}$~= 15~mJ/(mol K$^2$), which is strongly incompatible with the experiment. The aforementioned discrepancies are resolved by means of DFT+DMFT approach where strong electron correlations are treated more accurately in comparison to DFT+U method.
The obtained DFT+DMFT solution is metallic with strongly renormalized electronic structure near the Fermi energy. 
One should note that while there is a strong Stoner-like spitting for the Fe 3$d$ states with completely empty and completely occupied spin channels, the Re 5$d$ states are of Hubbard type with low and upper features below and above the Fermi level (Hubbard bands) and a narrow quasiparticle peak near the Fermi level. This peak means excitations that can enhance the effective mass which is inverse of the quasiparticle residue.  Therefore, the moderate mass enhancement of the electronic states leads to a renormalization of the linear electronic specific heat, $\gamma_{DFT+DMFT}$~= 74~mJ/(mol K$^2$), which is in a reasonable agreement with the experimental data and compatible with the magnetic properties of the \cucu~ compound.

\section{Conclusions}  To sum up, in the present paper we have studied effect of strong dynamical correlations on electronic and magnetic properties of quadruple perovskite \cucu. On the one hand, the DFT+DMFT calculations (taking these effects explicitly in account) agree partially with a static DFT+U in charge states of transition metal ions and magnetic moments. On the other hand, dynamical effects strongly affect electronic structure of the considered material and spoil the half-metallic state by closing a gap in the  minority spin channel, also providing considerable enhancement of electronic specific heat and renormalization of magnetic moment.  Moreover, correlation effects lead to formation of several intense peaks close to the Fermi level, which are absent in the DFT+U calculations. 
Such features can be related to Hubbard bands which are described in terms of incoherent states observed in optical transitions \cite{Raimondi,Irkhin}. Our results suggest that a proper account of electronic correlations is essential for quadruple perovskites and motivate further experimental spectroscopic studies. 

{\bf Acknowledgements.}
The authors are grateful to Zhehong Liu  for discussing experimental results on quadruple perovskites.
The DMFT calculations were supported by the Russian Science Foundation via project RSF 23-42-00069, while experiments by the National Natural Science Foundation of China (Grant No 12261131499).
The DFT calculations were performed  within the state assignment (the Ministry of Science and Higher Education of the Russian Federation, theme ``Quantum'' No. 122021000038-7).

\end{document}